\documentclass{svjour3}                     
\smartqed  
\usepackage{amssymb}
\begin{document}

\title{Quark confinement mechanism and the scale $\Lambda_{QCD}$
}


\author{Yu. P. Goncharov 
}


\institute{Yu. P. Goncharov \at
Theoretical Group, Experimental Physics Department, 
State Polytechnical University, 
Sankt-Petersburg 195251, Russia\\
              \email{ygonch@chern.hop.stu.neva.ru}           
}

\date{Received: date / Accepted: date}

\maketitle

\begin{abstract}
The confinement mechanism proposed earlier by the author is applied 
to problem of arising the so-called scale $\Lambda_{QCD}$ within the 
framework of QCD. The natural physical assumption consists of that 
$1/\Lambda_{QCD}\,\sim\,<r>$ where $<r>$ is a characteristic size of hadron
(radius of confinement). The above confinement mechanism allows us to calculate $<r>$ for mesons 
in terms of quark and gluonic degrees of freedom and this permits to conclude 
that $\Lambda_{QCD}$ should slightly change from hadron to hadron.

\keywords{Quantum chromodynamics \and Confinement
}

\end{abstract}

\section{Introduction}
As is known (see, e.g., \cite{{Hu},{Yn},{Du}}), within the framework of perturbative 
quantum chromodynamic (QCD) in asymptotical region of large momenta there 
appears some constant $\Lambda_{QCD}$ with dimension of mass, 
$\Lambda_{QCD}$ being not connected with arbitrary choice of the renormalization 
point. This is the so-called scale $\Lambda_{QCD}$ (in what follows we denote  
it by just $\Lambda$) and it cannot be determined 
from the theory in asymptotical region of large momenta and should arise 
in some way in the region of small momenta. The merits of case is contained in 
the relation between the effective strong coupling constant $\alpha_s$ and 
the momentum transfer $\sqrt{Q^2}$ 

$$\alpha_s=\alpha_s(Q^2)=\frac{12\pi}{(33-2n_f)\ln{(Q^2/\Lambda^2)}} 
\eqno(1)$$
 
while $n_f$ is number of quark flavours and (1) holds true at the momentum 
transfer $\sqrt{Q^2}\to\infty$. As is clear from (1), if $Q^2\to\Lambda^2$ then 
$\alpha_s\to\infty$ and traditional physical interpretation of it consists in 
that at distances of order $r\sim 1/\Lambda$ ($r$ is distance between quarks) 
interaction of quarks becomes very strong and provides the quark confinement so 
at distances greater than $1/\Lambda$ quarks are unobservable (see, e.g., 
\cite{{Hu},{Yn}}). It should be noted, however, that in fact the numerical 
value of $\Lambda$ depends on the situation where the relation (1) is 
used \cite{pdg}. Under the circumstances, appearance of the scale $\Lambda$ 
should be closely connected with the confinement mechanism and the latter 
should give an evaluation prescription of the mean radius of confinement $<r>$ 
which is in essence just a characteristic size of one or another hadron. 

It is known, however, at present no generally accepted quark confinement 
mechanism exists that would be capable to calculate a number of 
nonperturbative parameters characterizing mesons (masses, radii, decay 
constants and so on) 
appealing directly to quark and gluon degrees of freedom related to 
QCD-Lagrangian. At best there are a few scenarios (directly not connected to 
QCD-Lagrangian) of confinement that restrict themselves mainly to qualitative 
considerations with small possiblities of concrete calculation. In view of it 
in \cite{{Gon01},{Gon051},{Gon052}} a confinement mechanism has been proposed 
which was 
based on the {\em unique} family of compatible nonperturbative solutions for 
the Dirac-Yang-Mills system directly derived from QCD-Lagrangian. 
The word {\em unique} should be understood in the strict mathematical sense. 
Let us write down arbitrary SU(3)-Yang-Mills field in the form 
$A=A_\mu dx^\mu=A^a_\mu \lambda_adx^\mu$ ($\lambda_a$ are the 
known Gell-Mann matrices, $\mu=t,r,\vartheta,\varphi$, $a=1,...,8$ and we 
use the ordinary set of local spherical coordinates
$r,\vartheta,\varphi$ for spatial part of the flat Minkowski spacetime). 

In fact in \cite{{Gon01},{Gon051},{Gon052}} the following theorem was proved:

{\em The unique exact spherically symmetric (nonperturbative) solutions 
(depending only on $r$ and $r^{-1}$) of SU(3)-Yang-Mills equations in Minkowski spacetime 
consist of the family of form} 
$$ {\mathcal A}_{1t}\equiv A^3_t+\frac{1}{\sqrt{3}}A^8_t =-\frac{a_1}{r}+A_1 \>,
{\mathcal A}_{2t}\equiv -A^3_t+\frac{1}{\sqrt{3}}A^8_t=-\frac{a_2}{r}+A_2\>,$$
$${\mathcal A}_{3t}\equiv-\frac{2}{\sqrt{3}}A^8_t=\frac{a_1+a_2}{r}-(A_1+A_2)\>, $$
$$ {\mathcal A}_{1\varphi}\equiv A^3_\varphi+\frac{1}{\sqrt{3}}A^8_\varphi=
b_1r+B_1 \>,
{\mathcal A}_{2\varphi}\equiv -A^3_\varphi+\frac{1}{\sqrt{3}}A^8_\varphi=
b_2r+B_2\>,$$
$${\mathcal A}_{3\varphi}\equiv-\frac{2}{\sqrt{3}}A^8_\varphi=
-(b_1+b_2)r-(B_1+B_2)\> \eqno(2)$$

with the real constants $a_j, A_j, b_j, B_j$ parametrizing the family. 
Besides in \cite{{Gon01},{Gon051},{Gon052}} it was shown that the above unique 
confining solutions 
(2) satisfy the so-called Wilson confinement 
criterion \cite{{Wil},{Ban}}. Up to now 
nobody contested this result so if we want to describe interaction between 
quarks by spherically symmetric SU(3)-fields then they can be only the ones 
from the above theorem. On the other hand, the desirability of 
spherically symmetric (colour) interaction between quarks at all distances 
naturally follows from analysing the $p\bar{p}$-collisions (see, e.g., 
\cite{Per}) where one observes a Coulomb-like potential in events which 
can be identified with scattering quarks on each other, i.e., actually at small 
distances one observes the Coulomb-like part of solution (2). Under 
this situation, a natural assumption will be that the quark interaction remains 
spherically symmetric at large distances too but then, if trying to extend 
the Coulomb-like part to large distances in a spherically symmetric way, we 
shall inevitably come to the solution (2) in virtue of the above theorem.  

The aim of the present paper is to some extent to discuss a possible 
connection between the above confinement mechanism and the appearance of 
the scale $\Lambda$ in QCD. Section 2 contains a description of 
the confinement mechanism in question. 
Section 3 is devoted to how the mean radius of confinement could be 
computed within the framework under discussion while 
Section 4 is devoted to discussion and concluding remarks. Finally, 
Appendix A supplements the present section with a proof of the above uniqueness 
theorem in the case of SU(3)-Yang-Mills equations.

\section{Quark confinement mechanism}
The applications of the family (2) to the description of both the heavy 
quarkonia spectra \cite{{Gon031},{Gon032},{Gon04},{Gon08a}} and a number 
of properties of pions, kaons, $\eta$- and $\eta^\prime$-mesons 
\cite{{Gon06},{Gon07a},{Gon07b},{Gon08},{Gon08b},{Gon10}} 
showed that the confinement mechanism is qualitatively the same for both light 
mesons and heavy quarkonia. At this moment it can be decribed in the following 
way.

The next main physical reasons underlie linear confinement in the 
mechanism under discussion. The first one is that gluon exchange between 
quarks is realized with the propagator different from the photon-like one, and 
existence and form of such a propagator is a {\em direct} consequence of the 
unique confining 
nonperturbative solutions of the Yang-Mills equations 
\cite{{Gon01},{Gon051},{Gon052}}. The second reason is that, 
owing to the structure of the mentioned propagator, quarks mainly emit and 
interchange the soft gluons so the gluon condensate (a classical gluon field) 
between quarks basically consists of soft gluons (for more details 
see \cite{{Gon01},{Gon051},{Gon052}}) but, because of the fact that any gluon 
also emits gluons (still softer), the corresponding gluon concentrations 
rapidly become huge and form a linear confining magnetic colour field of 
enormous strengths, which leads to confinement of quarks. This is by virtue of 
the fact that just the magnetic part of the mentioned propagator is responsible 
for a larger portion of gluon concentrations at large distances since the 
magnetic part has stronger infrared singularities than the electric one. 
In the circumstances physically nonlinearity of the Yang-Mills equations 
effectively vanishes so the 
latter possess the unique nonperturbative confining solutions of the 
Abelian-like form (2) (with the values in Cartan subalgebra of SU(3)-Lie 
algebra) which describe the gluon condensate under 
consideration. Moreover, since the overwhelming 
majority of gluons is soft they cannot leave the hadron (meson) until some 
gluons obtain additional energy (due to an external reason) to rush out. So 
we also deal with the confinement of gluons.  

As has been repeatedly explained in 
\cite{{Gon031},{Gon032},{Gon04},{Gon08a},{Gon06},{Gon07a},{Gon07b},{Gon08},{Gon08b},{Gon10}}, 
parameters $A_{1,2}$ of 
solution (2) are inessential for physics in question and we can 
consider $A_1=A_2=0$. Obviously we have 
$\sum_{j=1}^{3}{\cal A}_{jt}=\sum_{j=1}^{3}{\cal A}_{j\varphi}=0$ which 
reflects the fact that for any matrix 
${\cal T}$ from SU(3)-Lie algebra we have ${\rm Tr}\,{\cal T}=0$. 
Also, as has been repeatedly discussed by us earlier (see, e. g., 
\cite{{Gon031},{Gon032},{Gon04},{Gon08a},{Gon06},{Gon07a},{Gon07b},{Gon08},{Gon08b},{Gon10}}), 
from the above form it is clear that 
the solution (2) is a configuration describing the electric Coulomb-like colour 
field (components $A^{3,8}_t$) and the magnetic colour field linear in $r$ 
(components $A^{3,8}_\varphi$) and we wrote down
the solution (2) in the combinations that are just 
needed further to insert into the Dirac equation (directly derived from 
QCD-Lagrangian) giving the interaction energy between two quarks in a meson 
and which will look as follows [after inserting solution (2)] 
$$i\partial_t\Psi\equiv  
i\pmatrix{\partial_t\Psi_1\cr \partial_t\Psi_2\cr \partial_t\Psi_3\cr}=
H\Psi\equiv
\pmatrix{H_1&0&0\cr 0&H_2&0\cr 0&0&H_3\cr}
\pmatrix{\Psi_1\cr\Psi_2\cr\Psi_3\cr}=
\pmatrix{H_1\Psi_1\cr H_2\Psi_2\cr H_3\Psi_3\cr}
                   \,,\eqno(3)$$
where Hamiltonian $H_j$ is 
$$H_j=\gamma^0\biggl[\mu_0-i\gamma^1\partial_r-i\gamma^2\frac{1}{r}
\left(\partial_\vartheta+\frac{1}{2}\gamma^1\gamma^2\right)- $$
$$i\gamma^3\frac{1}{r\sin{\vartheta}}
\left(\partial_\varphi+\frac{1}{2}\sin{\vartheta}\gamma^1\gamma^3
+\frac{1}{2}\cos{\vartheta}\gamma^2\gamma^3\right)\biggr]$$
$$-g\gamma^0\left(\gamma^0{\cal A}_{jt}+\gamma^3\frac{1}{r\sin{\vartheta}}
{\cal A}_{j\varphi}\right) \eqno(4)  $$                           
with the gauge coupling constant $g$, $\mu_0$ is a mass parameter and one 
should consider $\mu_0=m_{q_1}m_{q_2}/(m_{q_1}+m_{q_2})$ to be the reduced 
mass composed of the current masses $m_{q_1,q_2}$ of 
quarks forming a meson (quarkonium) while $\Psi=(\Psi_1, \Psi_2, \Psi_3)$ 
with the four-dimensional Dirac spinors $\Psi_j$ representing the $j$th colour 
component of the meson, so $\Psi$ may describe relative motion of two quarks 
in mesons. 

Additional considerations show that 
the {\em unique} modulo square integrable (nonperturbative) solutions of the 
Dirac equation (3) in the field (2) (i.e. {\em relativistic bound states}) 
are (with Pauli matrix $\sigma_1$, for more details see 
\cite{{Gon01},{Gon051},{Gon052}}) 
$$\Psi_j=e^{-i\omega_j t}\psi_j\equiv 
e^{-i\omega_j t}r^{-1}\pmatrix{F_{j1}(r)\Phi_j(\vartheta,\varphi)\cr\
F_{j2}(r)\sigma_1\Phi_j(\vartheta,\varphi)}\>,j=1,2,3\eqno(5)$$
with the 2D eigenspinor $\Phi_j=\pmatrix{\Phi_{j1}\cr\Phi_{j2}}$ of the
euclidean Dirac operator ${\cal D}_0$ on the unit sphere ${\Bbb S}^2$, while 
the coordinate $r$ stands for the distance between quarks. The explicit form of 
$\Phi_j$ is not needed here and can be found in 
\cite{{Gon031},{Gon032},{Gon04},{Gon08a},{Gon06},{Gon07a},{Gon07b},{Gon08},{Gon08b},{Gon10}}. 
One can only remark that spinors $\Phi_j$ form an orthonormal basis in 
$L_2^2({\Bbb S}^2)$ (in what follows we denote $L_2(F)$ 
the set of the modulo square integrable complex functions on any manifold $F$ 
furnished with an integration measure, then $L^n_2(F)$ will be the $n$-fold 
direct product of $L_2(F)$ endowed with the obvious scalar product).

 We can call the quantity $\omega_j$ 
relative energy of $j$th colour component of meson (while $\psi_j$ is wave 
function of a stationary state for $j$th colour component). Under this 
situation, if a meson is composed of quarks $q_{1,2}$ with 
different flavours then the energy spectrum of the meson will be given 
by $\epsilon=m_{q_1}+m_{q_2}+\omega$ with the current quark masses $m_{q_k}$ (
rest energies) of the corresponding quarks and we should put the interaction 
energy $\omega=\omega_j$ 
for any $j$ in virtue of Dirac equaion (3). On the other hand, 
$\omega_j$ is given by (for more details see 
Refs. \cite{{Gon01},{Gon051},{Gon052}})
$$\omega_j=\omega_j(n_j,l_j,\lambda_j)=$$ 
$$\scriptsize{
\frac{\Lambda_j g^2a_jb_j\pm(n_j+\alpha_j)
\sqrt{(n_j^2+2n_j\alpha_j+\Lambda_j^2)\mu_0^2+g^2b_j^2(n_j^2+2n_j\alpha_j)}}
{n_j^2+2n_j\alpha_j+\Lambda_j^2}
}
\>,\eqno(6)$$

where $a_3=-(a_1+a_2)$, $b_3=-(b_1+b_2)$, $B_3=-(B_1+B_2)$, 
$\Lambda_j=\lambda_j-gB_j$, $\alpha_j=\sqrt{\Lambda_j^2-g^2a_j^2}$, 
$n_j=0,1,2,...$, while $\lambda_j=\pm(l_j+1)$ are
the eigenvalues of euclidean Dirac operator ${\cal D}_0$ 
on unit sphere with $l_j=0,1,2,...$. 

In line with the above we should have the interaction energy $\omega=\omega_1=
\omega_2=\omega_3$ in energy spectrum $\epsilon=m_{q_1}+m_{q_2}+\omega$ for any 
meson (quarkonium) and this at once imposes two conditions on parameters 
$a_j,b_j,B_j$ when choosing some experimental value for $\epsilon$ at the 
given current quark masses $m_{q_1},m_{q_2}$. 

For example, for reference we shall give the radial parts of (5) 
at $n_j=0$ (the ground state) that are 
$$F_{j1}=C_jP_jr^{\alpha_j}e^{-\beta_jr}\left(1-
\frac{gb_j}{\beta_j}\right), P_j=gb_j+\beta_j, $$
$$F_{j2}=iC_jQ_jr^{\alpha_j}e^{-\beta_jr}\left(1+
\frac{gb_j}{\beta_j}\right), Q_j=\mu_0-\omega_j\eqno(7)$$
with $\beta_j=\sqrt{\mu_0^2-\omega_j^2+g^2b_j^2}$ while $C_j$ is determined 
from the normalization condition
$\int_0^\infty(|F_{j1}|^2+|F_{j2}|^2)dr=\frac{1}{3}$. 
Consequently, we shall gain that $\Psi_j\in L_2^{4}({\Bbb R}^3)$ at any 
$t\in{\Bbb R}$ and, as a result,
the solutions of (6) may describe ground state 
of mesons or quarkonia. The same holds true for wave functions (5) in general 
form so that (5) may describe {\em relativistic bound states} of mesons 
(quarkonia) with the energy (mass) spectrum $\epsilon=m_{q_1}+m_{q_2}+\omega$. 

\section{Meson spectroscopy and the scale $\Lambda_{QCD}$}

We would take a sin upon our soul if we did not try using the above {\em unique} 
family of {\em compatible} nonperturbative solutions for the Dirac-Yang-Mills 
system directly derived from QCD-Lagrangian for description of confinement 
since such an approach to confinement has good mathematical and physical grounds 
and, as a result, the approach is itself unique, nonperturbative and 
relativistic from the outset and it appeals immediately to quark and gluon 
degrees of freedom as should be from QCD-point of view. 

Having the off-the-shelf wave functions for mesons of the form (5) 
we can directly calculate $<r>$ in accordance with the standard 
quantum mechanics rules as $<r>=\sqrt{\int r^2\Psi^{\dag}\Psi d^3x}=
\sqrt{\sum\limits_{j=1}^3\int r^2\Psi^{\dag}_j\Psi_j d^3x}$ and the 
result will, e.g., for the case of ground state (7), be equal to (for more 
details see 
\cite{{Gon08a},{Gon06},{Gon07a},{Gon07b},{Gon08},{Gon08b},{Gon10}})

$$<r>=\sqrt{\sum\limits_{j=1}^3\frac{2\alpha^2_j+3\alpha_j+1}
{6\beta_j^2}}\>, \eqno(8) $$

and it is just the radius of meson (quarkonium) determined 
by the wave functions of (7) (at $n_j=0=l_j$) with respect to strong 
interaction, i.e., radius of confinement.  

As an illustration of (6) and (8), Tables I and II (taken from \cite{Gon08b}) 
contain the relevant numerical results for charged pions and kaons where it was 
accepted $m_u=2.25$ MeV, $m_d=5$ MeV, $m_s=107.5$ MeV when computing.

\begin{table*}
\caption{Gauge coupling constant, reduced mass $\mu_0$ and
parameters of the confining SU(3)-gluonic field for charged pions and kaons}
\label{t.1}
\begin{center}
\begin{tabular}{|c|c|c|c|c|c|c|c|c|}
\hline
\small Particle & \small $ g$ & \small $\mu_0$ (\small MeV) & \small $a_1$ 
& \small $a_2$ & \small $b_1$ (\small GeV) & \small $b_2$ (\small GeV) 
& \small $B_1$ & \small $B_2$ \\
\hline
\scriptsize $\pi^\pm$---$u\bar{d}$, $\bar{u}d$  
& \scriptsize 6.09131
& \scriptsize 1.55172
& \scriptsize 0.0473002
& \scriptsize 0.0118497
& \scriptsize 0.178915 
& \scriptsize -0.119290
& \scriptsize -0.230
& \scriptsize  0.230 \\
\hline
\scriptsize $K^\pm$---$u\bar{s}$, $\bar{u}s$  
& \scriptsize 5.30121
& \scriptsize 2.20387
& \scriptsize 0.167182
& \scriptsize -0.0557501
& \scriptsize 0.120150
& \scriptsize 0.131046
& \scriptsize -0.900
& \scriptsize  0.290 \\
\hline
\end{tabular}
\end{center}
\end{table*}

\begin{table*}
\caption{Theoretical and experimental masses and radii for charged pions and 
kaons}
\label{t.2}
\begin{center}
\begin{tabular}{|c|c|c|c|c|} 
\hline
\tiny Particle & \tiny Theoret. $\mu$ (MeV) &  \tiny Experim. $\mu$ (MeV) & 
\tiny Theoret. $<r>$ (fm)  & \tiny Experim. $<r>$ (fm)  \\
\hline
\scriptsize $\pi^\pm$---$u\bar{d}$, $\bar{u}d$   & \scriptsize $\mu= m_u+m_d+
\omega_j(0,0,1)= 139.570$ & \scriptsize 139.56995 & \scriptsize 0.673837 & 
\scriptsize 0.672 \\
\hline
\scriptsize $K^\pm$---$u\bar{s}$, $\bar{u}s$   & \scriptsize $\mu= m_u+m_s+
\omega_j(0,0,1)= 493.677$ & \scriptsize 493.677 & \scriptsize 0.544342 & 
\scriptsize 0.560 \\
\hline
\end{tabular}
\end{center}
\end{table*}

Then, as was said early in the paper, we should put $1/\Lambda\,\sim\, <r>$ 
with $<r>$ of (8) and it is clear that $\Lambda$ is a function of the 
parameters of the confining SU(3)-gluonic field (2) and the current quark 
masses, i.e., $\Lambda$ is expressed through gluonic and quark degrees of 
freedom, as should be according to the first principles of QCD. Also it is 
clear that because of $<r>$ slightly changes from hadron to hadron (see 
Table II) then $\Lambda$ should do it as well. It should be noted that crucial 
role in generating $<r>$ or $\Lambda$ belongs to the magnetic colour field of 
(2) (parameters $b_j$). Indeed, as follows from (8) at $|b_j|\to\infty$ we 
have $<r>\,\sim\, \sqrt{\sum\limits_{j=1}^3\frac{1}
{(g|b_j|)^2}}$, so in the strong magnetic colour field when $|b_j|\to\infty$, 
$<r>\to 0$, while the meson wave functions of (5) and (7) behave as 
$\Psi_j\,\sim\,e^{-g|b_j|r}$, i.e., just the magnetic colour field of (2) 
provides two quarks with confinement. Thus, our confinement mechanism gives 
an explanation of appearing the scale $\Lambda$ in QCD.

\section{Concluding remarks}

The results of present paper as well as the ones of 
\cite{{Gon031},{Gon032},{Gon04},{Gon08a},{Gon06},{Gon07a},{Gon07b},{Gon08},{Gon08b},{Gon10}} 
allow one to speak about that the 
confinement mechanism elaborated in \cite{{Gon01},{Gon051},{Gon052}}  
gives new possibilities for considering many old problems of hadronic 
(meson) physics (such as nonperturbative computation of decay constants, masses 
and radii of mesons \cite{{Gon06},{Gon07a},{Gon07b},{Gon08},{Gon08b},{Gon10}}, 
chiral symmetry breaking \cite{{Gon08b},{Gon10}} and 
so forth) from the first principles of QCD immediately appealing 
to the quark and gluonic degrees of freedom. This is possible because the 
given mechanism is based on the unique family of compatible 
nonperturbative solutions for the Dirac-Yang-Mills system directly derived from 
QCD-Lagrangian and, as a result, the approach is itself nonperturbative, 
relativistic from the outset, admits self-consistent nonrelativistic limit 
and may be employed for any meson (quarkonium). Under the circumstances the 
words {\em quark and gluonic degrees of freedom} make exact sense: gluons come 
forward in the form of bosonic condensate described by parameters $a_j$, 
$b_j$, $B_j$ from the unique exact solution (2) of the Yang-Mills equations 
while quarks are represented by their current masses $m_q$.

Finally, one should say that the unique confining solutions similar to (2) 
exist for all semisimple and non-semisimple compact Lie groups, in particular, 
for SU($N$) with $N\ge2$ and 
U($N$) with $N\ge1$ \cite{{Gon01}}. Explicit form of solutions, 
e.g., for SU($N$) with $N=2,4$ can be found in \cite{Gon052} but it 
should be emphasized that components linear in $r$ always represent the 
magnetic (colour) field in all the mentioned solutions. Especially the case 
U(1)-group is interesting which corresponds to usual electrodynamics. 
Under this situation, as was pointed out in \cite{{Gon051},{Gon052}} there is 
an interesting possibility of 
indirect experimental verification of the confinement mechanism under 
discussion. Indeed the confining solutions 
of Maxwell equations for classical electrodynamics point out 
the confinement phase could be in electrodynamics as well. Though 
there exist no elementary charged particles generating a constant magnetic 
field linear in $r$, the distance from particle, after all, if it could 
generate this elecromagnetic field configuration in laboratory then one might 
study motion of the charged particles in that field. The confining properties 
of the mentioned field should be displayed at classical level too but the exact 
behaviour of particles in this field requires certain analysis of the corresponding 
classical equations of motion. Such a program has been recently realized in 
\cite{GF10}. Motion of a charged (classical) particle was studied in the 
field representing magnetic part of the mentioned solution of Maxwell equations 
and it was shown that one deals with the full classical confinement of the 
charged particle in such a field: under any initial conditions the particle 
motion is accomplished within a finite region of space so that the particle 
trajectory is near magnetic field lines while the latter are compact manifolds 
(circles). Those results might be useful in thermonuclear plasma physics 
(for more details see \cite{GF10}). 

\section*{Appendix A}
The facts adduced here have been obained in Refs. \cite{{Gon051},{Gon052}} and 
we concisely give them only for completeness of discussion. 

To specify the question, let us note that in general the Yang-Mills equations 
on a manifold $M$ can be written as
$$d\ast F= g(\ast F\wedge A - A\wedge\ast F) \>,\eqno(\mathrm A.1)$$ 
where a gluonic field $A=A_\mu dx^\mu=
A^a_\mu \lambda_adx^\mu$ [$\lambda_a$ are the 
known Gell-Mann matrices, $\mu=t,r,\vartheta,\varphi$ (in the case of 
spherical coordinates), $a=1,...,8$], 
the curvature matrix (field strentgh)
$F=dA+gA\wedge A= F^a_{\mu\nu}\lambda_adx^\mu\wedge dx^\nu$ with exterior 
differential $d$ and the Cartan's (exterior) product $\wedge$, while $\ast$ 
means the Hodge star operator conforming to a metric on manifold under 
consideration, $g$ is a gauge coupling constant.

The most important case of $M$ is Minkowski spacetime and we 
are interested in the confining solutions $A$ of the SU(3)-Yang-Mills 
equations. The confining solutions were defined in Ref. \cite{Gon01} as the 
spherically symmetric solutions of the Yang-Mills 
equations (1) containing only the components of the 
SU($3$)-field which are Coulomb-like or linear in $r$. Additionally 
we impose the Lorentz condition on the sought solutions. 
The latter condition is necessary for 
quantizing the gauge fields consistently within the framework of perturbation 
theory (see, e. g. Ref. \cite{Ryd85}), so we should impose the given condition 
that can be written
in the form ${\rm div}(A)=0$, where the divergence of the Lie algebra valued
1-form $A=A_\mu dx^\mu=A^a_\mu \lambda_adx^\mu$ is defined by the relation 
(see, e. g., Ref. \cite{Bes87})
$${\mathrm{div}}\, {A}=\ast(d\ast{A})=
\frac{1}{\sqrt{\delta}}\partial_\mu(\sqrt{\delta}g^{\mu\nu}
A_\nu)\>.\eqno(\mathrm A.2)$$
It should be emphasized that, from the physical point of view, the Lorentz 
condition reflects the fact of transversality for gluons that arise as quanta 
of SU(3)-Yang-Mills field when quantizing the latter (see, e. g., 
Ref. \cite{Ryd85}).

We shall use the Hodge star operator action on the 
basis differential 2-forms on Minkowski spacetime with local 
coordinates $t, r, \vartheta, \varphi$ in the form
$$\ast(dt\wedge dr)=-r^2\sin\vartheta d\vartheta\wedge d\varphi\>,
\ast(dt\wedge d\vartheta)=\sin\vartheta dr\wedge d\varphi\>,$$
$$\ast(dt\wedge d\varphi)=-\frac{1}{\sin\vartheta}dr\wedge d\vartheta\>,
\ast(dr\wedge d\vartheta)=\sin\vartheta dt\wedge d\varphi\>,$$
$$\ast(dr\wedge d\varphi)=-\frac{1}{\sin\vartheta}dt\wedge d\vartheta\>,
\ast(d\vartheta\wedge d\varphi)=\frac{1}{r^2\sin\vartheta}dt\wedge dr\>,
\eqno(\mathrm A.3)$$
so that on 2-forms $\ast^2=-1$. More details about the Hodge star operator can 
be found in \cite{Bes87}. 

The most general ansatz for a spherically symmetric solution is 
$A=A_t(r)dt+A_r(r)dr+A_\vartheta(r)d\vartheta+A_\varphi(r)d\varphi$. 
But then the Lorentz 
condition (A.2) for the given ansatz gives rise to 
$$\sin{\vartheta}\partial_r(r^2A_r)+
\partial_\vartheta(\sin{\vartheta}A_\vartheta)=0,$$
which yields $A_r=\frac{C}{r^2}-
\frac{\cot{\vartheta}}{r^2}\int A_\vartheta(r)dr$ with a constant matrix $C$. 
But the confining solutions should be spherically symmetric and contain only 
the components which are Coulomb-like or linear in $r$, so one should put 
$C=A_\vartheta(r)=0$. Consequently, 
the ansatz $A=A_t(r)dt+A_\varphi(r)d\varphi$ is the most general 
spherically symmetric one. 

For the latter ansatz we have $F=dA+gA\wedge A=-\partial_rA_tdt\wedge dr+
\partial_rA_\varphi dr\wedge d\varphi+g[A_t,A_\varphi]dt\wedge d\varphi$, 
where $[\cdot,\cdot]$ signifies matrix commutator.
 
Then, according to (A.3), we obtain 
$$\ast F= (r^2\sin{\vartheta})\partial_rA_td\vartheta\wedge d\varphi-
\frac{1}{\sin{\vartheta}}\partial_rA_\varphi dt\wedge d\vartheta-$$ 
$$\frac{g}{\sin{\vartheta}}[A_t,A_\varphi]dr\wedge d\vartheta\>,
\eqno(\mathrm A.4)$$ 
which entails 
$$d\ast F= 
\sin{\vartheta}\partial_r(r^2\partial_rA_t)\,dr\wedge d\vartheta\wedge d\varphi+
\frac{1}{\sin{\vartheta}}\partial_r^2A_\varphi\, dt\wedge dr\wedge d\vartheta
\>,\eqno(\mathrm A.5)$$
while 
$$\ast F\wedge A - A\wedge\ast F=$$
$$\biggl(r^2\sin{\vartheta}[\partial_rA_t,A_t]-
\frac{1}{\sin{\vartheta}}[\partial_rA_\varphi,A_\varphi]\biggr)
\,dt\wedge d\vartheta\wedge d\varphi$$
$$-\frac{g}{\sin{\vartheta}}\biggl([[A_t,A_\varphi],A_t]\,
dt\wedge dr\wedge d\vartheta$$
$$+[[A_t,A_\varphi],A_\varphi]\,
dr\wedge d\vartheta\wedge d\varphi\biggr)
\>.\eqno(\mathrm A.6)$$
Under the circumstances the Yang-Mills equations (A.1) are 
tantamount to the conditions 
$$\partial_r(r^2\partial_rA_t)=-
\frac{g^2}{\sin^2{\vartheta}}[[A_t,A_\varphi],A_\varphi],\eqno(\mathrm A.7)$$
$$\partial_r^2A_\varphi=-
{g^2}[[A_t,A_\varphi],A_t],\eqno(\mathrm A.8)$$
$$r^2\sin{\vartheta}[\partial_rA_t,A_t]-
\frac{1}{\sin{\vartheta}}[\partial_rA_\varphi,A_\varphi]=0.
\eqno(\mathrm A.9)$$
The key equation is (A.7) because the matrices $A_t, A_{\varphi}$ depend on 
merely $r$ and (A.7) can be satisfied only if the matrices 
$A_t=A_t^a\lambda_a$ and $A_\varphi=A_{\varphi}^a\lambda_a$ belong to the 
so-called Cartan subalgebra of the SU(3)-Lie algebra. Let us remind that, by definition, 
a Cartan subalgebra is a maximal abelian subalgebra in 
the corresponding Lie algebra, i. e., the commutator for any two matrices of 
the Cartan subalgebra is equal to zero (see, e.g., Ref. \cite{Bar}). For 
SU(3)-Lie algebra the conforming Cartan subalgebra is generated by the 
Gell-Mann matrices $\lambda_3, \lambda_8$ which are
$$\lambda_3=\pmatrix{1&0&0\cr 0&-1&0\cr 0&0&0\cr}\,,  
  \lambda_8={1\over\sqrt3}\pmatrix{1&0&0\cr 0&1&0\cr 
                   0&0&-2\cr}\,.\eqno(\mathrm A.10)$$
Under the situation we should have $A_t=A_t^3\lambda_3+A_t^8\lambda_8$ and 
$A_\varphi=A_{\varphi}^3\lambda_3+A_{\varphi}^8\lambda_8$, then 
$[A_t,A_\varphi]=0$ and we obtain 
$$\partial_r(r^2\partial_rA_t)=0, \partial_r^2A_\varphi=0, 
\eqno(\mathrm A.11)$$
while (A.9) is identically satisfied and (A.11) gives rise to the solution (2) with 
real constants $a_j, A_j, b_j, B_j$
parametrizing the solution which proves the uniqueness theorem of Section 1  
for the SU(3) Yang-Mills equations. 

The more explicit form of (2) is 
$$A^3_t = [(a_2-a_1)/r+A_1-A_2]/2,\>$$
$$A^8_t =[A_1+A_2-(a_1+a_2)/r]\sqrt{3}/2\>,$$
$$ A^3_\varphi = [(b_1-b_2)r+B_1-B_2]/2,$$
$$ A^8_\varphi= [(b_1+b_2)r+B_1+B_2]\sqrt{3}/2\>.\eqno(\mathrm A.12)$$

Clearly, the obtained results may be extended over all SU($N$)-groups with 
$N\ge2$ and even 
over all semisimple compact Lie groups since for them the corresponding Lie 
algebras possess just the only Cartan subalgebra. Also we can talk about the 
compact non-semisimple groups, for example, U($N$). In the latter case 
additionally to Cartan subalgebra we have centrum consisting from the matrices 
of the form $\alpha I_N$ ($I_N$ is the unit matrix $N\times N$) with arbitrary 
constant $\alpha$. 

The most relevant physical cases are of course U(1)- and SU(3)-ones 
(QED and QCD). In particular, the U(1)-case allows us to build the classical 
model of confinement (see Ref. \cite{GF10}). 

At last, it should also be noted that the 
nontrivial confining solutions obtained exist at any gauge coupling constant 
$g$, i. e. they are essentially {\em nonperturbative} ones.



\end{document}